\documentclass[showpacs,superscriptaddress,preprint]{revtex4}
\usepackage{amsmath,graphicx}

\begin{document}

\title{Percolation on hyperbolic lattices}
\author{Seung Ki Baek}
\email{garuda@tp.umu.se}
\affiliation{Department of Theoretical Physics, Ume{\aa} University, 901 87 Ume{\aa}, Sweden}
\author{Petter Minnhagen}
\email{Petter.Minnhagen@physics.umu.se}
\affiliation{Department of Theoretical Physics, Ume{\aa } University, 901 87
Ume{\aa }, Sweden}
\author{Beom Jun Kim}
\email{beomjun@skku.edu}
\affiliation{Department of Physics, BK21 Physics Research Division, Sungkyunkwan
University, Suwon 440-746, Korea}

\begin{abstract}
The percolation transitions
on hyperbolic lattices are investigated numerically using finite-size
scaling methods.
The existence of two distinct percolation thresholds
is verified. At the lower threshold, an unbounded cluster appears and reaches
from the middle to the boundary. This transition is of the same type and
has the same finite-size scaling properties as the corresponding transition
for the Cayley tree.
At the upper threshold, on the other hand, a single unbounded
cluster forms which overwhelms all the others and occupies a finite
fraction of the volume as well as of the boundary connections.
The finite-size scaling properties for this upper threshold are different
from those of the Cayley tree and two of the critical exponents are obtained.
The results suggest that the percolation transition for the  hyperbolic
lattices forms a universality class of its own. 

\end{abstract}

\pacs{64.60.ah, 02.40.Ky, 05.70.Fh}

\maketitle

\section{Introduction}
\label{sec:intro}

Geometry is often crucial in physical phenomena, since dimensionality and
topological defects determine the properties of phase transitions. Progress
in complex networks is an example where geometrical features are
paramount to the physical properties (for a review, see, e.g., Ref.~\cite
{doro}). This is also reflected in ongoing research interests in the
nontrivial lattice structures such as fractal lattices~\cite{fractals} and
Apollonian networks~\cite{apollo}. Another important area is
the hyperbolic geometry where even the familiar standard physical
models like electronic spins and Brownian motions exhibit novel
behaviors~\cite{heptagonal}. Such studies are interesting not only from a
theoretical viewpoint but also from a potential real applicability related
to the rapid development of fabrication of devices and structures in
nanoscales~\cite{nano}. In this work, we focus on the percolation problem
and investigate the percolation phase transition for negatively curved
hyperbolic lattices. Percolation on hyperbolic lattices has so far been of
predominantly mathematical interest and several interesting mathematical
results have been reported, including the existence of an intermediate phase
with infinitely many unbounded 
clusters~\cite{benjamini,schon,lyons}. More specifically, there exist in
general two critical thresholds: unbounded clusters are
formed at the first threshold,
while unbounded clusters are merged to become
one unique unbounded cluster at the second threshold.
In case of the usual flat lattices, which have vanishing surface-volume
ratios in infinite-volume limit, the two thresholds coincide,
and the theory of such percolation transitions is
well-developed~\cite{stauffer}. On the other hand, the properties of the
percolation transitions for hyperbolic lattices still remain to be further
clarified. In particular, these lattices are not homogeneous
due to the presence of a boundary and as a consequence, the critical
properties may differ from the mean-field-type transitions which were
discussed in Ref.~\cite{schon2}. In this paper, we investigate the
characteristic features of the
thresholds and the corresponding phases by using various statistical
measures like the number of boundary points connected to the
middle, the ratio of the first and second largest clusters, and the cluster
size distribution. We use finite-size scaling methods to obtain the
critical properties, together with the Newman-Ziff
algorithm~\cite{newman-ziff}.

This paper is organized as follows: In Sec.~\ref{sec:percs}, we introduce
alternative manifestations of the percolation thresholds and show that
all of them coincide in an ordinary square lattice. In
Sec.~\ref{sec:hyperb}, we explain the notion of a hyperbolic lattice and
the appearance of two distinct percolation transitions.
We start with the Bethe lattice and the standard mean-field results. Next the
Cayley tree is introduced which is the simplest model with two thresholds. 
 This model is used as a benchmark when discussing the characteristic
features of percolation in hyperbolic lattices. The efficiency of the
finite-size scaling methods is also illustrated for this simpler model. Then
we present the results for the general hyperbolic structures which contain
loops. We summarize our results in Sec.~\ref{sec:summary}.

\section{Various manifestations of percolation}
\label{sec:percs}

\begin{figure}[tbp]
\includegraphics[width=0.45\textwidth]{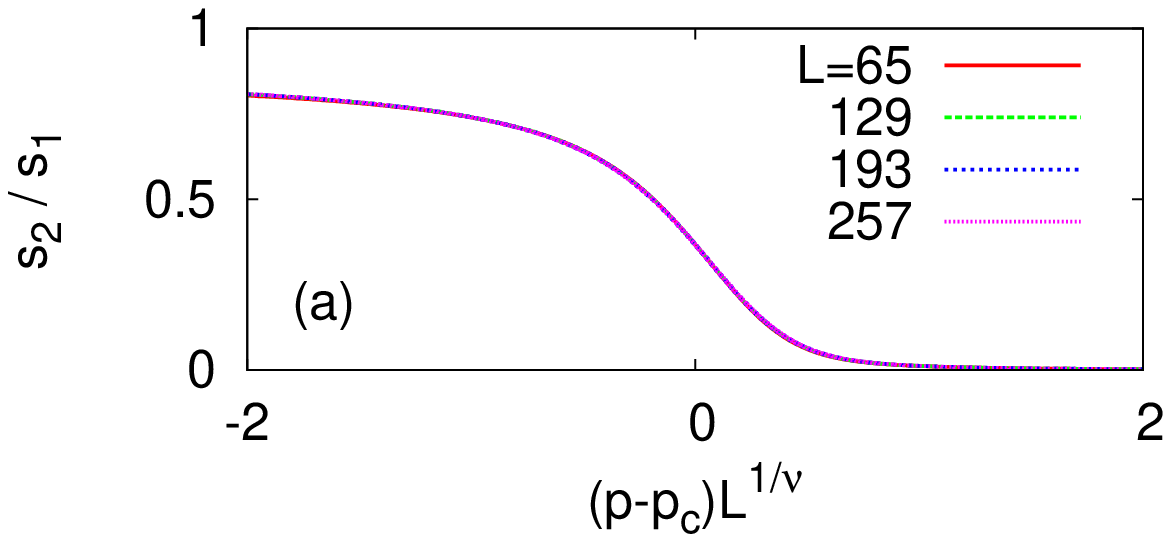} 
\includegraphics[width=0.45\textwidth]{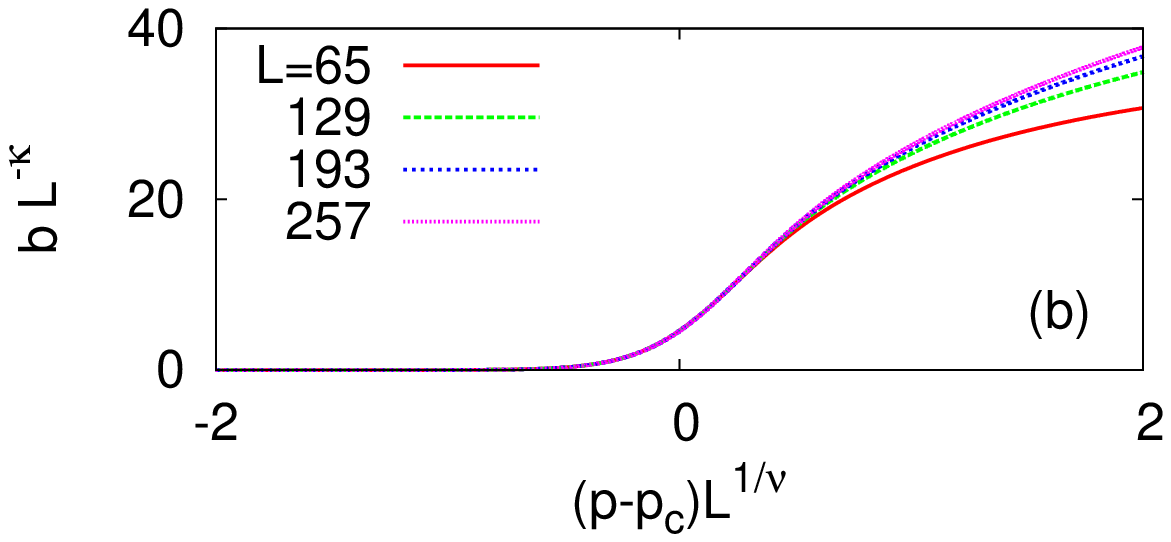} 
\caption{(Color online) Scaling plots for regular square lattices, using
(a) the ratio between the largest and the second largest cluster
sizes, and (b) the number of
boundary points connected to the midpoint of the lattice.
The two-dimensional percolation scaling exponent $\protect\nu=4/3$ and
$\kappa = \nu / (1+\nu) = 4/7$ are used for scaling collapse with
$p_u=p_c=1/2$.
The average is taken over $10^6$ independent realizations for each plot.
}
\label{fig:square}
\end{figure}

There are various possible manifestations of percolation.
We here describe the critical thresholds corresponding to them as well as
their relationship.
The percolation threshold is usually defined as the occupation probability
$p$ above which a cluster is formed which occupies a finite fraction of the
system.
Let us first consider an $L\times L$ square lattice.
In this case it is well-known that $p_c=1/2$ for bond percolation
\cite{stauffer}.
Alternatively, we may consider the ratio between the largest cluster
size, $s_1$, and the second largest one, $s_2$~\cite{zen}. 
This measure is based on the fact that above the transition threshold,
the second
largest cluster becomes negligible with respect to the largest one, and
accordingly, as soon as $p$ exceeds a critical value from below, $s_2/s_1$
vanishes in the large lattice limit.
Since it implies that one large unique cluster dominates the whole
system, we call this threshold $p_u$. 
Figure~\ref{fig:square}(a) illustrates that the ensemble average of
$s_2/s_1$ scales quite well with sizes for the square lattice.
This finite-size scaling supports
the well-known result that $p_u=p_c=1/2$ and the critical index
$\nu=4/3$. We will use this type of finite-size scaling methods throughout
the paper.

An alternative
manifestation of percolation 
is the number $b$ of boundary points connected to the middle
of the lattice via occupied bonds. We will also use this concept of midpoint
percolation throughout the paper.
Around the threshold of the midpoint percolation, which we will call $p_m$, a
cluster penetrates the whole system and one expects a finite-size scaling
of the form
\begin{equation}
b=L^{\kappa }f[(p-p_m)L^{1/\nu }]  \label{eq:bscaling}
\end{equation}
with $\kappa = \nu/(1+\nu) = 4/7 \approx 0.57$~\cite{sapoval,forthcoming}.
Again, Fig.~\ref{fig:square}(b) shows an excellent scaling collapse
with $\kappa = 4/7$ and $p_m=p_c=1/2$.
Consequently, the finite size scaling of $b$ gives a practical alternative
way of investigating percolation properties.
Yet another measure, based on the same midpoint percolation concept, is the
fraction of the boundary points connected to the middle, $b/4L$, which
becomes finite when $p$ gets above $p_b$.
This measure will be also frequently used.
For the planar lattices all these thresholds coincide so that we
have only one critical threshold
$p_c=p_u=p_m=p_b=1/2$. (The observable quantities used in the present paper are
listed in Table~\ref{table:thresholds}.) The crucial point in the present
context is that this equality does not hold for hyperbolic lattices.
\begin{table}
\caption{Percolation thresholds and quantities used for their numerical
detections. Here $s_i$ represents the $i$th largest cluster size, $b$ is
the number of boundary points connected to the middle of the lattice, and
$B$ is the total number of boundary points.}
\begin{tabular}{|c|c|c|}\hline\hline
Threshold & Meaning & Observable\\ \hline
$p_u$ & The unbounded cluster becomes unique. & $s_2/s_1$\\
$p_m$ & The boundary is connected to the middle. & $b$\\
$p_b$ & A finite fraction of the boundary is connected to the middle. &
$b/B$\\ \hline\hline
\end{tabular}
\label{table:thresholds}
\end{table}

\section{Percolation in hyperbolic lattices}
\label{sec:hyperb}

\begin{figure}
\includegraphics[width=0.3\textwidth]{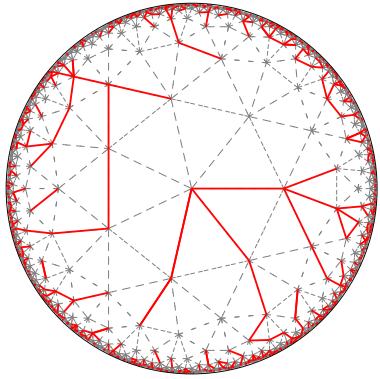} 
\caption{(Color online) Poincar\'e disk representation of the hyperbolic
lattice $\{m,n\} = \{3,7\}$ [seven ($n=7$) triangles ($m=3$) meet at every
vertex] with the total number of layers $l=6$. Solid lines show occupied
bonds in a realization with the occupation probability of $p=0.2$.}
\label{fig:3_7visual}
\end{figure}

Suppose that $n$ regular $m$-gons meet at every vertex, which we represent
by a Schl\"afli symbol $\{m,n\}$~\cite{coxeter1954}. The resulting lattices
are flat if we take $\{m,n\}=\{3,6\},\{4,4\},\{6,3\}$, which together
constitute the two-dimensional (2D) percolation universality class~\cite
{stauffer}. If $(m-2)(n-2)>4$, on the other hand, the resulting lattice is
known to have a constant negative Gaussian curvature~\cite{coxeter1997},
making a hyperbolic surface. Figure~\ref{fig:3_7visual} shows a
hyperbolic lattice represented as $\{3,7\}$, where seven regular triangles
meet at each vertex. Note that
all of the triangles in this figure are congruent with respect to the metric
used in this projection on the Poincar\'e disk~\cite{anderson}.
As shown in Fig.~\ref
{fig:3_7visual}, we construct the lattice in a concentric way so that the
origin of the disk becomes the zeroth layer and its seven nearest neighbors
constitute the first layer. Likewise, the second layer surrounding the first
layer is composed of 21 vertices, and so on up to the $l$th layer [$l=6$
in Fig.~\ref{fig:3_7visual}].
The system is said to have a level $l$, and its size is given by
$N(l) = 1 + 7
\sum_{j=1}^{l}[c_+^j - c_-^j]/\sqrt{5}$ with $c_\pm = (3 \pm \sqrt{5})/2$.
For example, $N(l=2) = 1 + 7 \frac{(c_+ - c_-)}{\sqrt{5}} +
7 \frac{(c_+^2 - c_-^2)}{\sqrt{5}} = 1 + 7 + 21 = 29$.
This formula shows that the number of lattice points increases exponentially
with a distance from the origin $O$, yielding a nonvanishing surface-volume
ratio in the limit $N\rightarrow \infty$. In other words, the total number of
boundary points $B$ is proportional to the system size $N$.
The sizes of other lattices with different Schl\"afli symbols are listed in
Table~\ref{table:sizes}.

Since a $d$-dimensional hypercube
with a volume $v$ has the surface-volume ratio $\propto v^{-1/d}$, a
hyperbolic lattice is usually called infinite-dimensional. In graph theory,
an object with a nonvanishing surface-volume ratio is called
\emph{nonamenable}~\cite{benjamini,lyons}.
The exponential increase of lattice points as a function of the distance from the middle also constitute a practical challenge since the
range of possible sizes is limited when using the finite-size scaling method.
For this reason, we present numeric values only up to the second digit in this
work.

\begin{table}
\caption{System sizes for various structures, $\{m,n\}$, as a function of
level $l$. Each lattice is grown up from a single midpoint in the zeroth
layer.}
\begin{tabular}{|c|c|}\hline\hline
$\{m,n\}$ & $N(l)$ \\ \hline
$\{3,7\}$ & $1 + \frac{7}{\sqrt{5}} \sum_{j=1}^{l}[(\frac{3+\sqrt{5}}{2})^j -
(\frac{3-\sqrt{5}}{2})^j]$ \\
$\{4,5\}$ & $1 + \frac{5}{\sqrt{3}} \sum_{j=1}^{l}[(2+\sqrt{3})^j -
(2-\sqrt{3})^j]$ \\
$\{5,5\}$ & $1 + \sqrt{5} \sum_{j=1}^{l}[(\frac{7+3\sqrt{5}}{2})^j -
(\frac{7-3\sqrt{5}}{2})^j]$ \\
$\{6,4\}$ & $1 + 2\sqrt{2} \sum_{j=1}^{l}[(3+2\sqrt{2})^j -
(3-2\sqrt{2})^j]$ \\
$\{7,3\}$ & $1 + \frac{15}{\sqrt{5}} \sum_{j=1}^{l}[(\frac{3+\sqrt{5}}{2})^j -
(\frac{3-\sqrt{5}}{2})^j]$ \\
$\{\infty,3\}$ & $1 + 3\sum_{j=1}^{l} 2^{j-1}$\\ \hline\hline
\end{tabular}
\label{table:sizes}
\end{table}

\subsection{Bethe lattice}
\label{sec:bethe}

The binary Bethe lattice is the infinite binary tree where all the lattice
points are equivalent~\cite{soderberg}. This means that the Bethe lattice lacks
boundary points. Consequently, it belongs to the
amenable class. This is in contrast to the Cayley tree (discussed in
the following section) which includes the boundary even in the large lattice
limit and hence is an
example of a nonamenable graph. The percolation for the Bethe lattice is
exactly solvable and the solution corresponds to the standard mean-field
theory~\cite{stauffer}. This standard mean-field theory describes the
percolation transition for $d$-dimensional Euclidean lattices provided $d\geq
6$~\cite{stauffer}. However, it has limited applicability in the context of
nonamenable graphs.

The critical threshold $p_c$ is well-known since the early percolation
theory formulated in the gelation process~\cite{flory}. The point
is that percolation on a Bethe lattice can be treated as the Galton-Watson
branching process~\cite{grimmett1989}.
We pick up an arbitrary point as a root, and the set of all the points
reached from it by $i$ bonds is called the $i$th generation (the term
{\em generation} will be used in trees, instead of {\em layer}).
Let us denote $w$ as the extinction probability that the branching process
from the root is ended at some finite generation of the tree.
For such a process,
each bond to the next generation should be either unoccupied with a
probability $1-p$, or occupied but eventually terminated with the
probability $pw$. Since each vertex has two bonds to the next generation,
the sum of those probabilities has to be squared and $w$ satisfies a
self-consistency equation, $w = (1-p+pw)^2 $, yielding
\begin{equation*}
w = \left\{ 
\begin{array}{lcr}
1 & \mbox{for} & 0 \leq p<1/2, \\ 
(1/p-1)^2 & \mbox{for} & 1/2 \leq p \leq 1.
\end{array}
\right.
\end{equation*}
When $p > 1/2$, the extinction probability is less than unity.
From the percolation viewpoint, every vertex has one successor on average
at $p=1/2$, and accordingly, the cluster from the root vertex can be
extended indefinitely. Consequently, the bond
percolation of the Bethe lattice has $p_c = 1/2$, at which unbounded
clusters may be formed.
For a general Bethe lattice denoted as $\{\infty,n\}$, this generalizes into 
$p_c = 1/(n-1)$.

An important point in the present context is that amenable graphs only have
one percolation threshold. Thus for Bethe lattice the threshold $p_m$ [where a
cluster from the midpoint (which in this case is any point because all
points are equivalent) reaches a point arbitrarily far away]
and $p_c$ [the threshold where the unbounded clusters contain a finite
fraction of the whole system] are equal, i.e., $p_m=p_c$ due to the
equivalence between points. Also the three
planar lattices ($\{4,4\}$, $\{3,6\}$, $\{6,3\}$)
are amenable and consequently
have only one threshold (compare with Sec.~\ref{sec:percs}).
As will be discussed in
the following, nonamenable graphs such as hyperbolic lattices have two
distinct percolation thresholds.

\subsection{Cayley tree, $\{\infty,3\}$}
\label{sec:tree}

\begin{figure}[tbp]
\includegraphics[width=0.45\textwidth]{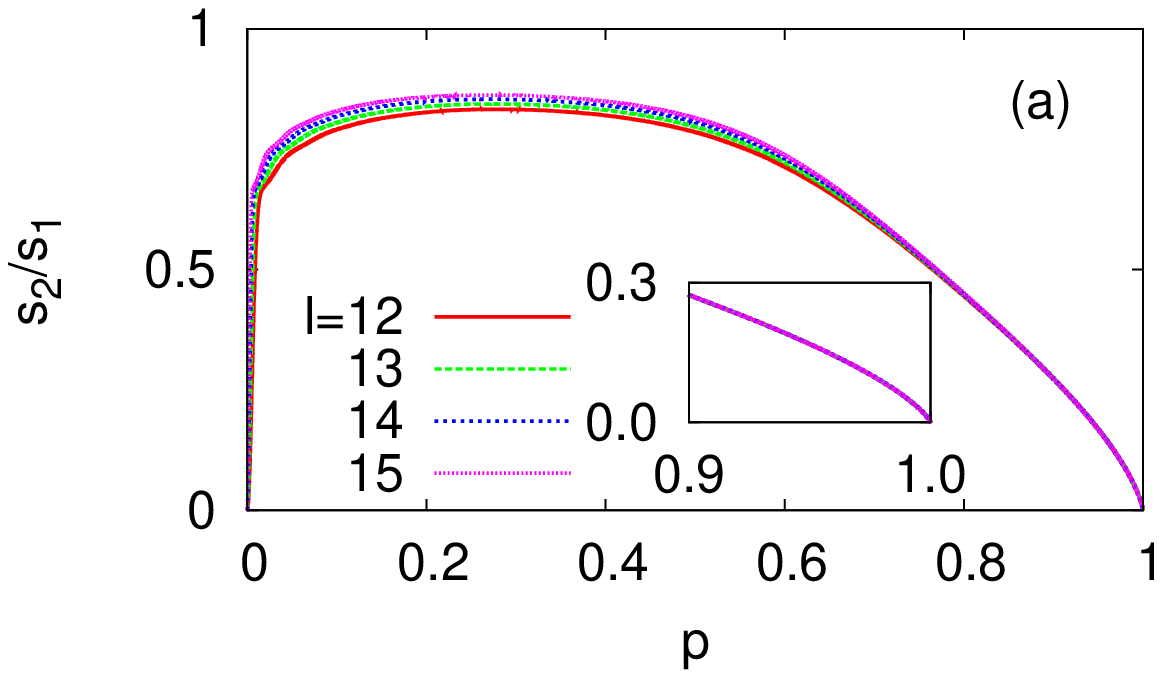} 
\includegraphics[width=0.45\textwidth]{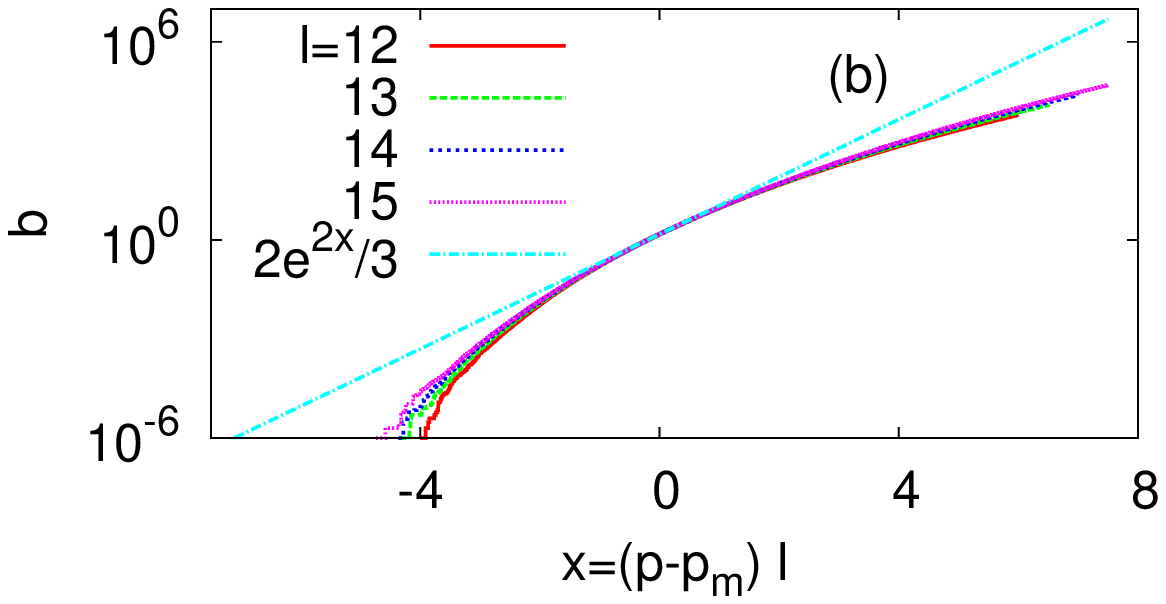} 
\includegraphics[width=0.45\textwidth]{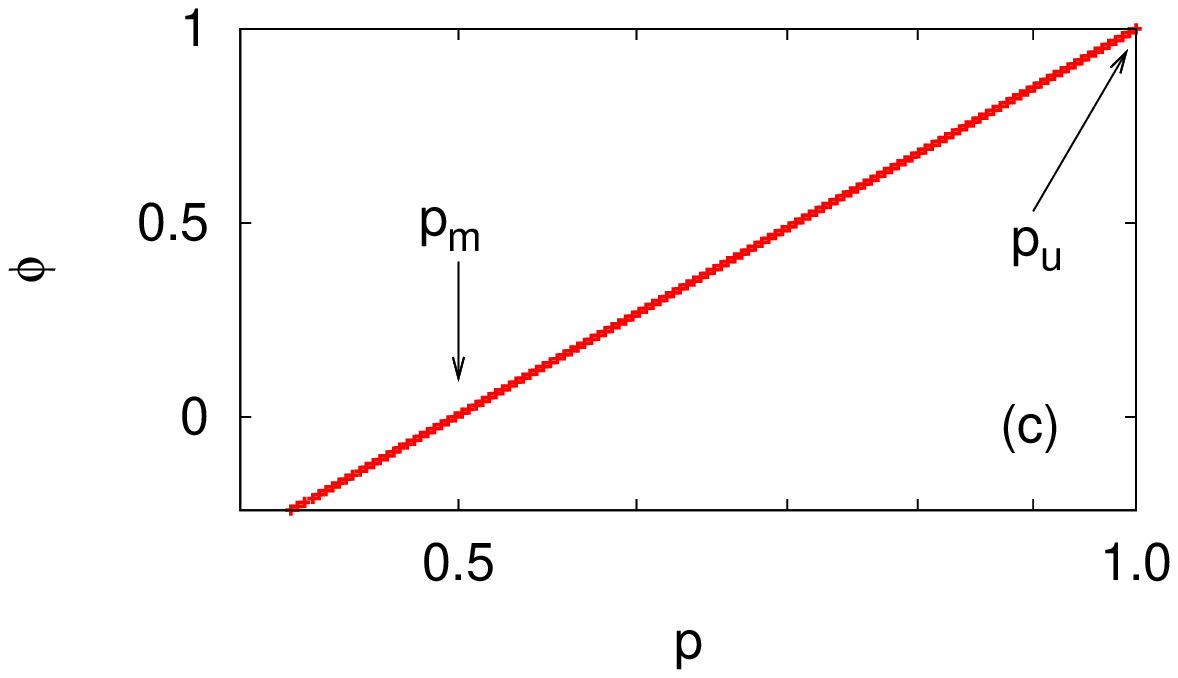} 
\includegraphics[width=0.45\textwidth]{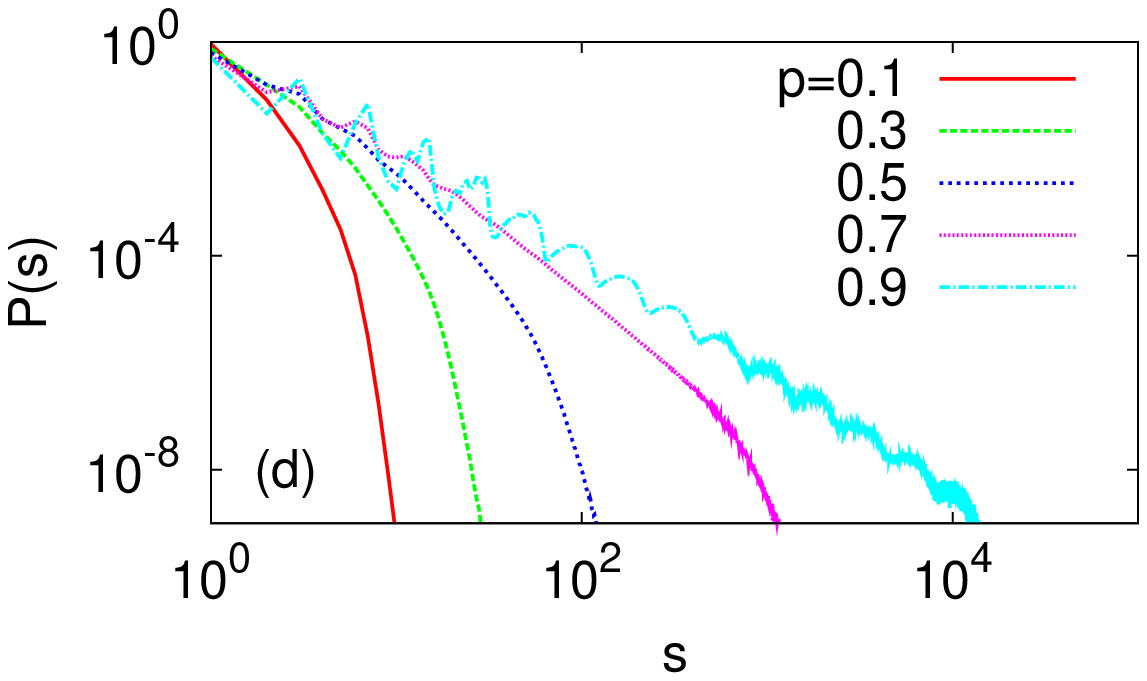} 
\caption{(Color online) Numerical results in Cayley trees, averaged over
$10^6$ realizations. (a)
The ratio between $s_2$ and $s_1$. Inset: A closer look near $p=1$, where
one can hardly see any size dependency. (b) Scaling plot by
Eq.~(\ref{eq:bscaling2}). The scaling function is well-described
by $\frac{3}{2} \exp[2l (p-p_m)]$ near $p_m=0.50$.
(c) $\phi$ as a function of $p$, provided that $b(p) \sim
B^{\phi}$.
The horizontal axis is in the log scale, confirming $\phi =
1+\log_2 ~p$.
(d) Cluster size distribution at various $p$ with $l=15$.
It approaches to $P(s) \sim
s^{-2}$ as $p \rightarrow 1$, while finiteness of the system adds an
exponential cutoff.
}
\label{fig:cayley}
\end{figure}

The Cayley tree is a tree grown from the root vertex up to the $l$th
generation where the root vertex is identified as the middle of the lattice.
This is an example of a nonamenable graph.  While it has sometimes been
presumed that the Bethe lattice is an adequate limiting case of the Cayley
tree, more recent studies suggest that the Cayley tree in the limit of $l
\rightarrow \infty$ has different critical properties from those of the
Bethe lattice~\cite{doro}.
Note that vertices are not equivalent for the Cayley tree
so that one can clearly define which
generation a vertex belongs to. The branching-process argument
is again applicable and the cluster from the root vertex reaches the
bottom of the tree at $p_m = 1/2$. On the other hand, the uniqueness
threshold is located at $p_u=1$~\cite{schon},
since $s_2/s_1$ remains finite at any $p<1$ [Fig.~\ref{fig:cayley}(a)].
The vanishing of $s_2/s_1$ as $p_u=1$ is approached can be obtained as follows:
The total number of bonds in the tree with a level $l$ is
$K=2^{l+1}-2$. Suppose that precisely one bond is broken on average. The
occupation probability corresponding to this is $p=(K-1)/K$. Suppose further
that the broken bond connects the $i$th and $(i+1)$th generations. The
probability to select this bond is given by
\begin{equation*}
P(i) = \frac{2^{i+1}}{2^{l+1}-2}.
\end{equation*}
Breaking one bond creates precisely two clusters.
The smaller one is below the broken bond and has a size of $s_2 =
2^{l-i}-1$. The size of the larger one is consequently $s_1 = N -
s_2 = 2^{l+1}-2^{l-i}$. The expectation value of the ratio $s_2/s_1$ is
obtained as follows:
\begin{eqnarray}
\left<\frac{s_2}{s_1}\right>
&=& \displaystyle \sum_{i=0}^{l-1}\frac{s_2}{s_1} P(i)\nonumber\\
&=& \displaystyle \sum_{i=0}^{l-1}\frac{2^{l-i}-1}{2^{l+1}-2^{l-i}}
\frac{2^{i+1}}{2^{l+1}-2}\nonumber\\
&=& \displaystyle \sum_{i=0}^{l-1}\frac{2^{l-i}-1}{2^{l+1}-2^{l-i}}
\frac{2^{i+1}}{2^{l+1}-2}\nonumber\\
&\approx& \displaystyle \frac{1}{2^{l+1}} \sum_{i=0}^{l-1}\frac{1}{1-2^{-i-1}}
\nonumber\\
&\approx& \displaystyle \frac{l}{2^{l+1}} \nonumber
\end{eqnarray}
Since $p=(K-1)/K$, one can express this result directly in terms of $p$. By
using the connection $1-p = K^{-1} \approx 2^{-l-1}$, one obtains
\begin{equation*}
\left<\frac{s_2}{s_1}\right> \propto -(1-p) \log (1-p).
\end{equation*}
Thus the ratio $s_2/s_1$ vanishes as $-(1-p) \log (1-p)$ as the threshold
$p_u=1$ is approached from below. This is illustrated in
Fig.~\ref{fig:cayley}(a): The Cayley tree has two thresholds and
$p_m<p_u$. 

The midpoint percolation quantity $b$ instead possesses the scaling form 
\begin{equation}
b= l^{\kappa} f[(p-p_m) l^{1/\nu}],  \label{eq:bscaling2}
\end{equation}
with $\kappa=0$ and $\nu=1$. This scaling form can be derived as follows:
The number of boundary points which is reached from the midpoint for a given
value $p$ is $b=3p \times (2p)^{l-1}$ from which it follows that  
\begin{equation}
b = \frac{3}{2} \exp[\log(2p)l] \approx \frac{3}{2}
\exp[2l(p-p_m)]. \label{eq:exact}
\end{equation}
The validity of this finite-size scaling is demonstrated in
[Fig.~\ref{fig:cayley}(b)] together with the exact scaling form given by
Eq.~(\ref{eq:exact}).

One may also note that since $l \approx \log_2 ~N$, it follows that $b \sim
(2p)^l = N p^l \approx N p^{\log_2 ~N} = N^{1+\log_2 ~p}$. This means that $b$
scales as $N^{\phi}$ (or equivalently as $B^{\phi}$) at any value of
$0<p\leq 1$ with a $p$-dependent exponent $\phi = 1+\log_2 ~p$
[Fig.~\ref{fig:cayley}(c)].
A direct consequence of this is that $b/B$
eventually goes to zero at every $p<1$ in the limit of infinite $N$. Thus $b/B$
is discontinuous at $p_b=1$ for $l=\infty$. This is in contrast to the ratio
$s_2/s_1$ which goes continuously to zero at $p_u=1$.

One can also obtain the limiting behavior of the cluster size distribution
$P(s)$ by removing one bond. In this case $P(s)~ds$ is the probability of
finding a cluster with a size in the interval $[s,s+ds]$ and $P(i)~di$ is the
probabililty of the single bond to be broken between generations $i$ and
$(i+1)$.
This latter probability is also equal to the probability of finding a second
largest cluster of size $s_2\propto s^{L-i}$. Thus $P(s_2)ds_2 = P(i)di$ from which follows that
\begin{equation*}
P(s_2) = P(i) \left| \frac{di}{ds_2} \right| \propto s_2^{-2},
\end{equation*}
where the last proportionality is obtained from $s_{2}\propto s^{L-i}$. This
suggests that the size distribution of clusters $P(s)$ (excluding the
largest one) should approach the form $P(s) \propto s^{-\tau}$ where $\tau
\rightarrow 2$ as $p \rightarrow 1$.
Since the generation $i$ in a tree is actually a discrete variable it follows
that the possible sizes of $s_2$ are also discrete and this discreteness
becomes noticeable as $p \rightarrow 1$. This is illustrated in
Fig.~\ref{fig:cayley}(d) which shows the approach to $P(s)\propto s^{-2}$ as
well as the log-periodic oscillations caused by the discreteness.

A more intuitive explanation for $P(s)$ may be gained by mapping the
percolation problem in a Cayley tree to a branching process, namely, the
formation of family trees~\cite{baek}:
Let us consider the family-size distribution in the $z$-ary tree ($z \equiv
n-1)$ at the $k$th generation,
with the net birth rate $\lambda = \log(zp)$ as each
family grows by $(zp)^k$. When a bond is broken in a tree graph, the
top of this detached branch is interpreted as the first ancestor of a new
family, and we describe this top vertex as an immigrant, whose number is
proportional to
the existing population size $N_k$. If the number of immigrants is written
as $\zeta N_k$, the birth and immigration should yield a constant growth of
population, $\zeta + \lambda = \log ~z$, as we know $N_{k+1} = z N_k$.
According to Ref.~\cite{baek}, the family size distribution at the $k$-th
generation exhibits a power law with an exponent 
$\tau^{\prime }= 2 + \zeta/\lambda = 2 + \log(1/p)/\log(zp)$ as $k
\rightarrow \infty$. Since the volume of a given tree is proportional to its
surface, it seems plausible that $\tau \approx \tau^{\prime }$ for the cluster
size distribution, $P(s)$. Note that $\tau'=2$ for $p=1$ which is also the limiting result for
the Cayley tree. This gives a hand-waving argument suggesting that the size
distribution form $P(s) \propto s^{-\tau}$ could be valid also for some
range of $p$ below one.

\subsection{Heptagonal lattice, $\{7,3\}$}

\begin{figure}[tbp]
\includegraphics[width=0.45\textwidth]{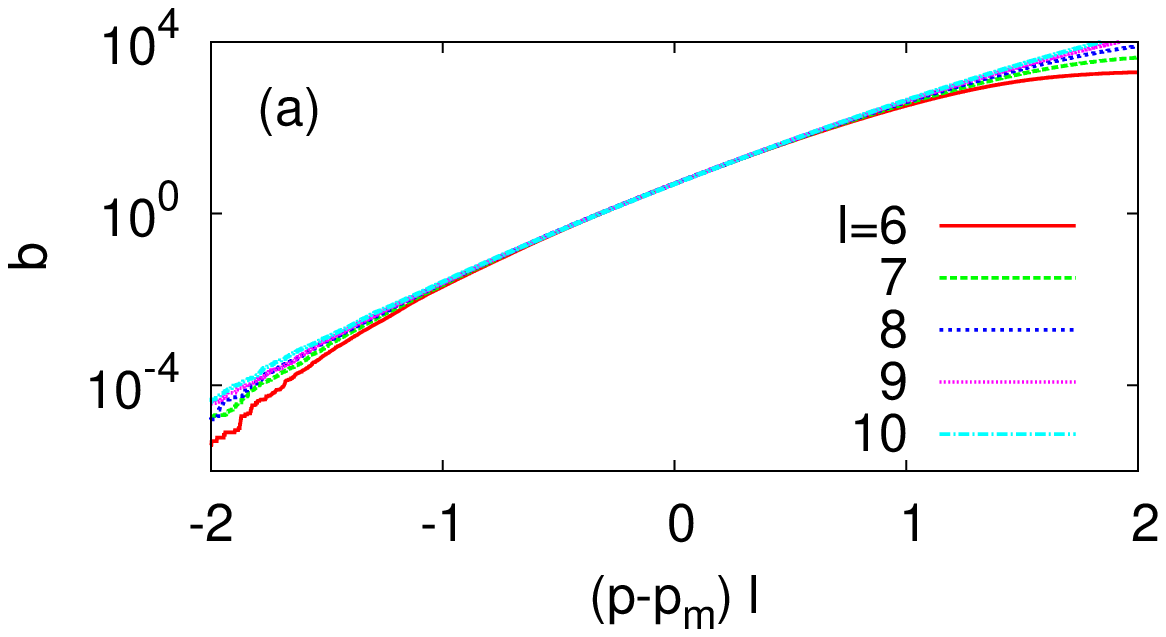} 
\includegraphics[width=0.45\textwidth]{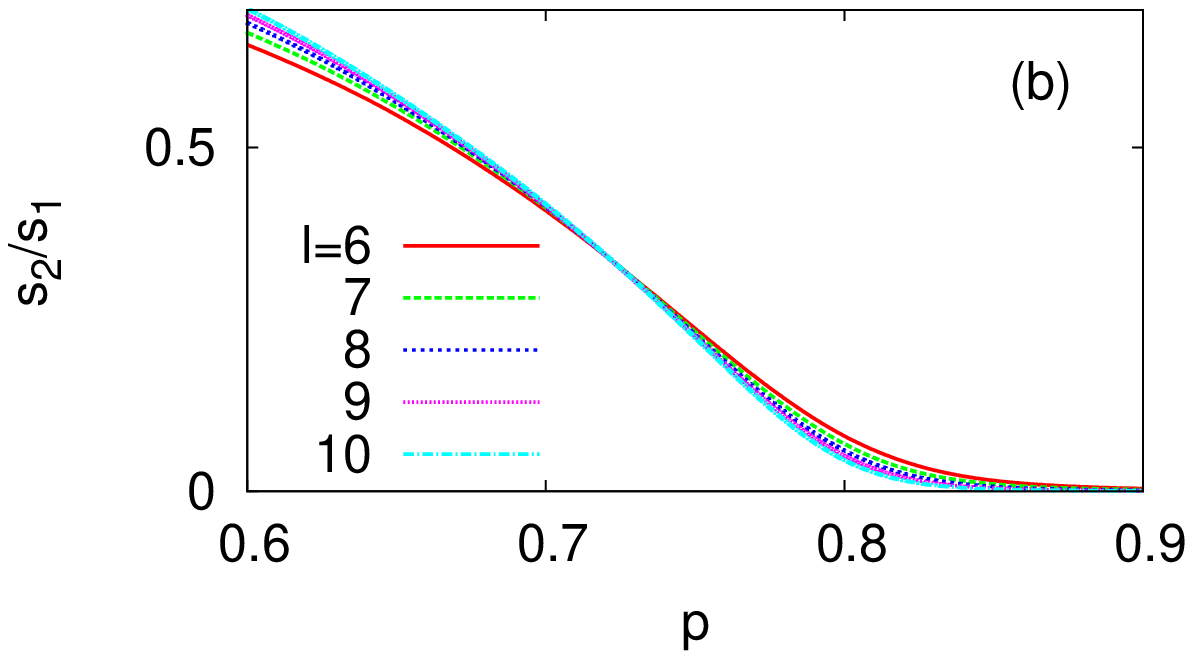} 
\includegraphics[width=0.45\textwidth]{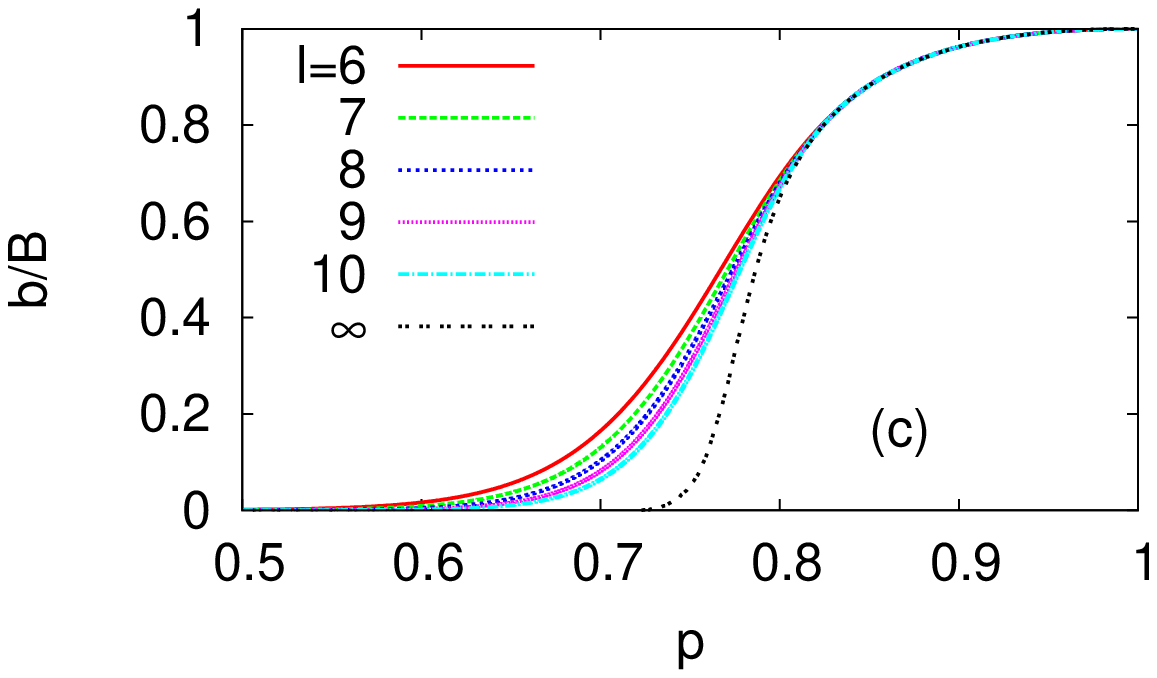} 
\includegraphics[width=0.45\textwidth]{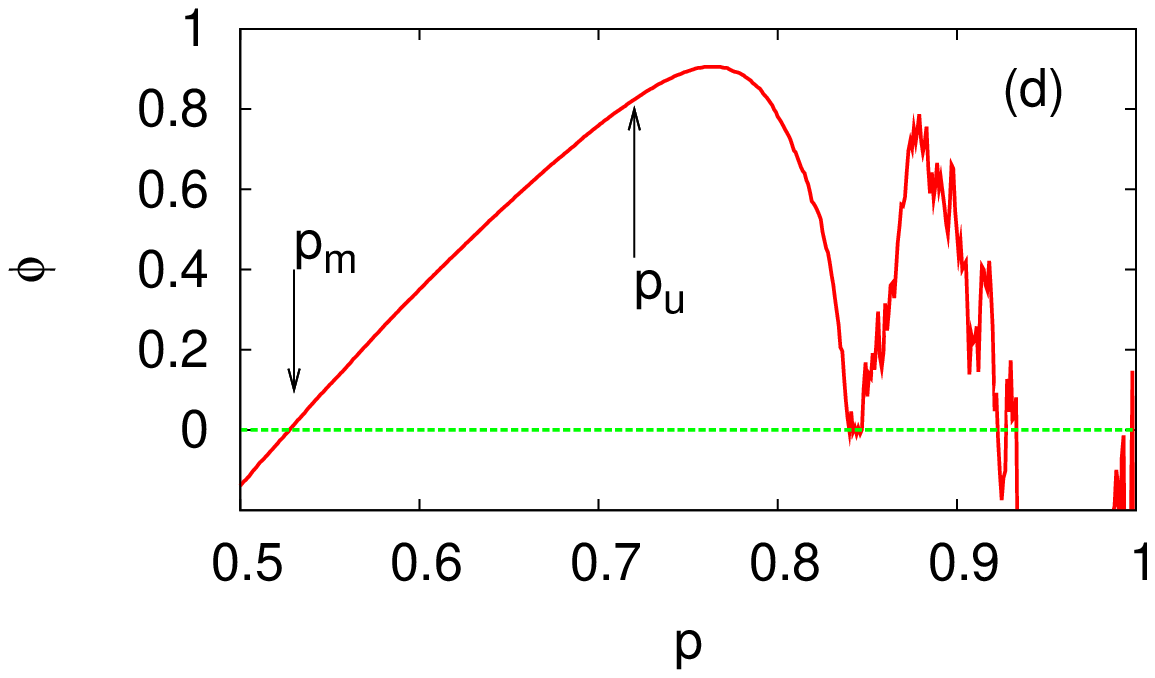} 
\caption{(Color online) Numerical results for heptagonal lattices,
$\{7,3\}$, averaged over $10^6$ realizations. (a)
Measurement of $b$ yields a consistent result with the tree case.
(b) The ratio between two largest cluster sizes $s_2/s_1$
gives $p_u \approx 0.72$, and (c) extrapolating $b/B$ gives $p_b
\approx 0.72$, supporting $p_b = p_u$. (d) The exponent $\phi$ as a
function of $p$ shows a clear difference from the tree case [see
Fig.~\ref{fig:cayley}(c), for comparison]. At $p \gtrsim 0.84$, $\phi$ largely
fluctuates as it is hard to determine the size dependency
from the numerical data.}
\label{fig:7_3}
\end{figure}

Now we consider a heptagonal lattice, denoted as $\{m,n\} = \{7,3\}$.
Since the probability to find a loop is roughly $O(p^m)$, we expect the
results obtained for a Cayley tree will remain valid to some extent.
The lower threshold is determined from the finite-size scaling of $b$ and
Eq.~(\ref{eq:bscaling2}) as shown in Fig.~\ref{fig:7_3}(a).
This determines the value of $p_m \approx 0.53$. We note that this is rather
close to the exact result for the Cayley tree $1/(n-1) = 1/2$, and shows the
same scaling form as in Eq.~(\ref{eq:bscaling2}). This suggests that the
loops have only a small effect on the percolation properties at small values
of $p$, as is also suggested by the actual structure of the clusters
illustrated in Fig.~\ref{fig:3_7visual}.

The upper threshold is determined from the finite-size scaling of $s_2/s_1$
which gives $p_u \approx 0.72$.
According to a dual-lattice argument~\cite{hunt}, a percolation threshold
for a regular lattice $\{m,n\}$ is predicted approximately as $m/(m+n)$.
Applying this argument somewhat {\it ad hoc} to the present case yields
$7/(7+3) = 0.7$, which is in fact fairly close to the actual value,
$p_u\approx 0.72$.

\begin{figure}[tbp]
\includegraphics[width=0.45\textwidth]{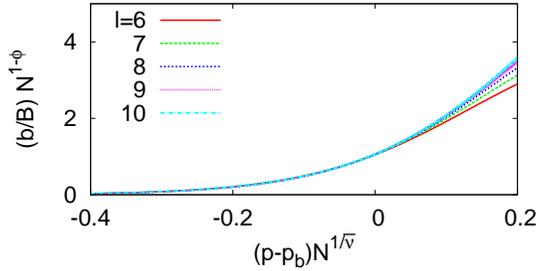} 
\caption{(Color online) Scaling plot of $b/B$ in $\{7,3\}$ using
Eq.~(\ref{eq:7_3scaling2}), with $p_b=0.72$, $\phi=0.82$, and
$1/\bar\nu=0.12$.}
\label{fig:7_3scale2}
\end{figure}

Alternatively, the upper threshold can be determined from $b/B$ by assuming
that the size scaling form is the same as for the Cayley tree,
\[b/B \sim c_1 N^{\phi-1} + c_2,\]
and extrapolating the numerical results from $l=6,\ldots,11$ to the infinite-size limit
[Fig.~\ref{fig:7_3}(c)]. This extrapolation gives $|b/B| \lesssim O(10^{-3})$
below $p<0.72$ and becomes positive finite above that,
suggesting $p_b \approx 0.72$. It thus suggests $p_b = p_u$
within our numerical accuracy. This corresponding scaling exponent $\phi(p)$ is
plotted in Fig.~\ref{fig:7_3}(d). While $\phi(p)$ is an increasing function
of $p$ in the Cayley tree, it is a convex function in $\{7,3\}$. The
crucial difference is that $\phi$ is still less than one at $p=p_b$.
In order to study the transition at $p_b$, we use the finite-size scaling form:
\begin{equation}
b/B \sim N^{\phi-1} f[(p-p_b)N^{1/\bar\nu}].
\label{eq:7_3scaling2}
\end{equation}
The scaling collapse at $p_b=0.72$ determines the critical indices to 
$\phi \approx 0.82$ and $\bar\nu \approx 0.12$ (Fig.~\ref{fig:7_3scale2}).

\begin{figure}[tbp]
\includegraphics[width=0.45\textwidth]{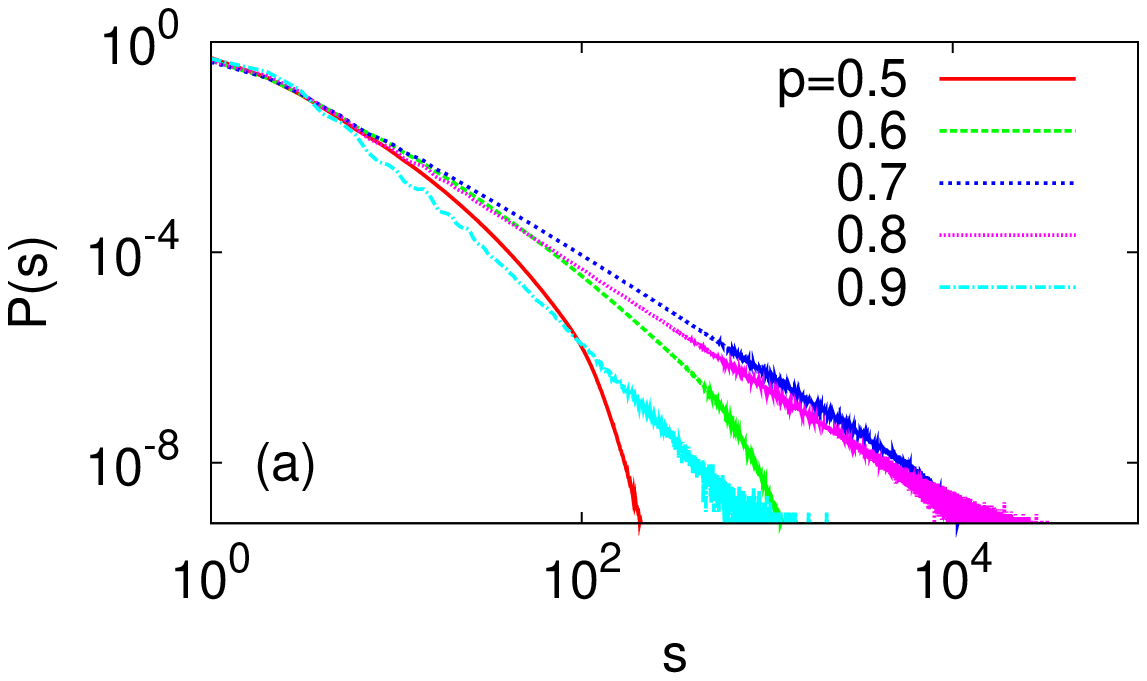} 
\includegraphics[width=0.25\textwidth]{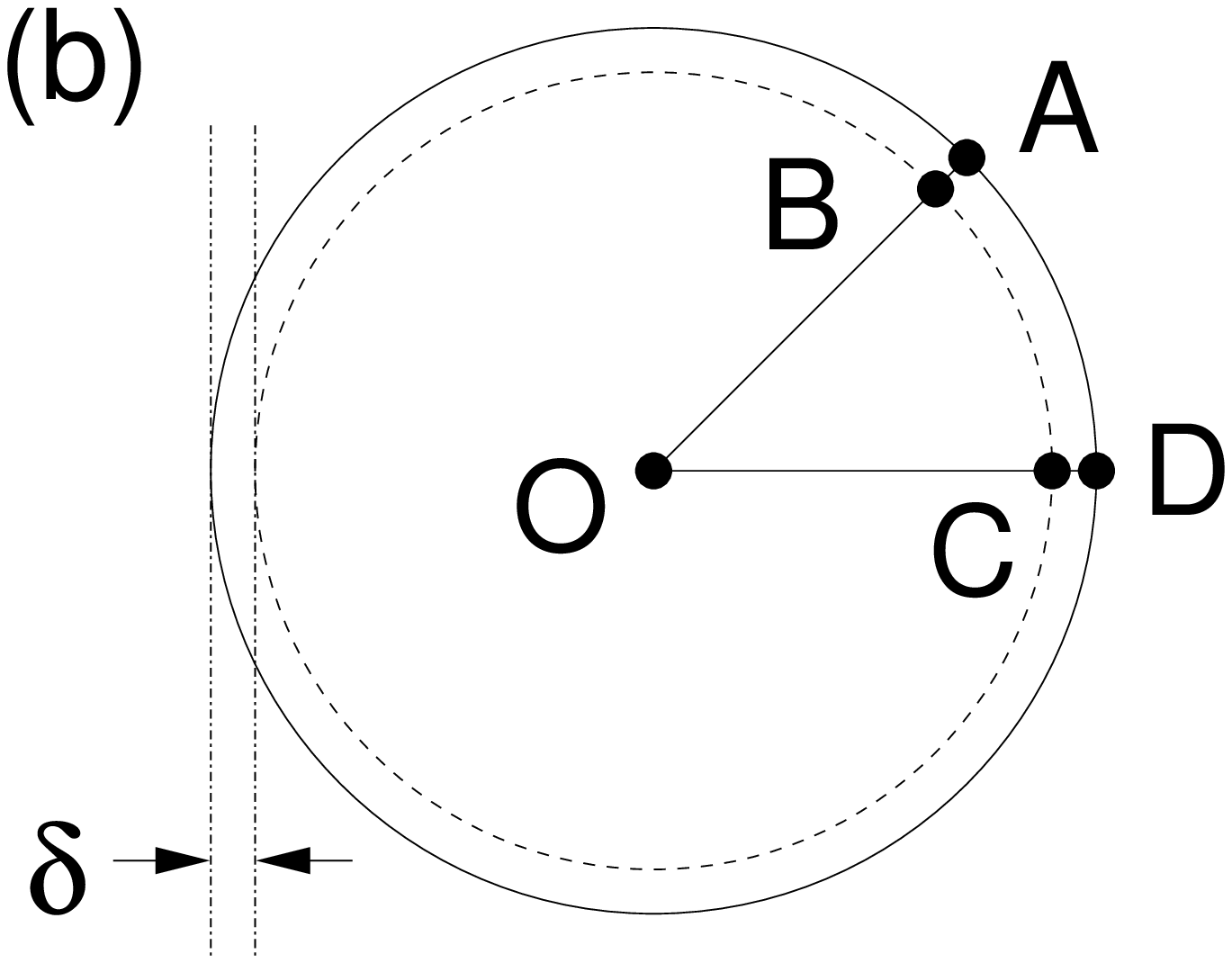} 
\caption{(Color online) (a) Cluster size distribution, $P(s)$, at various
$p$ in $\{7,3\}$. It shows a power-law form also above $p_u$. (b) A
schematic view of the Poincar\'e disk centered at $O$. The lattice is
constructed up to the dashed circle, which is apart from the circumference
of the disk by $\overline{AB}=\overline{CD}=\protect\delta$ in the Euclidean
measure.}
\label{fig:7_3csize}
\end{figure}

Another crucial difference from the Cayley tree is the appearance of a
supercritical region at $p>p_u$.
We find that the hyperbolic lattices in this region display a power-law
behavior in the cluster size distribution [Fig.~\ref{fig:7_3csize}(a)].
A hand-waving argument for this behavior goes as follows:
The probability of finding a cluster with a certain size is dominated by its
surface area, since connections through the surface have to be cut to isolate
this cluster from the surrounding. It is thus believed that $P(s) \propto
\exp[-\eta(p) s^{1-1/d}]$ in a $d$-dimensional lattice with some $p$-dependent
constant $\eta(p)$~\cite{stauffer,grimmett1989}. For a hyperbolic lattice
with $d=\infty$, one may well expect $P(s) \propto e^{-s}$. In case of our
nonamenable setting, however, most large-sized clusters are facing the
outmost boundary of the lattice, where the outward connections are already
absent. Suppose a cluster contained in a fan-shape $OBC$ on the Poincar\'e
disk, where the lattice is constructed up to the dashed line in
Fig.~\ref{fig:7_3csize}(b). If $ \overline{AB} = \overline{CD} = \delta$ in
the Euclidean measure, the hyperbolic length of the arc
$\overset{\frown}{BC}$ connecting $B$ and $C$ is of an order of
$\delta^{-1}$. Since the cluster size is closely related to the surface
length in a hyperbolic lattice, which is proportional to
$\delta^{-1}$~\cite{anderson}, we may say that $s \propto
\delta^{-1}$.
At the same time, the hyperbolic length that one should cut
out to isolate this cluster is roughly the length of the geodesic
connecting $B$ and $C$, which grows as $\log~\delta^{-1}$.
Note that $\overset{\frown}{BC}$ needs not be considered
here since it is a part of the lattice boundary,
and this makes the fundamental difference from the exponential decay.
In other words, the number of bonds cut for hyperbolic lattices
is not proportional to the size $ s $, but only to $\log ~s$. This gives the
cluster size distribution in a
power-law form as $P(s) \sim \exp[-\eta(p) \log ~s] = s^{-\eta(p)}$. It is also 
possible to infer that $\eta(p)$ should be an increasing function of $p$, as clusters
are merged to the largest one in the supercritical phase.

\subsection{Comparison with $\{4,5\}$ and $\{3,7\}$}

\begin{figure}[tbp]
\includegraphics[width=0.45\textwidth]{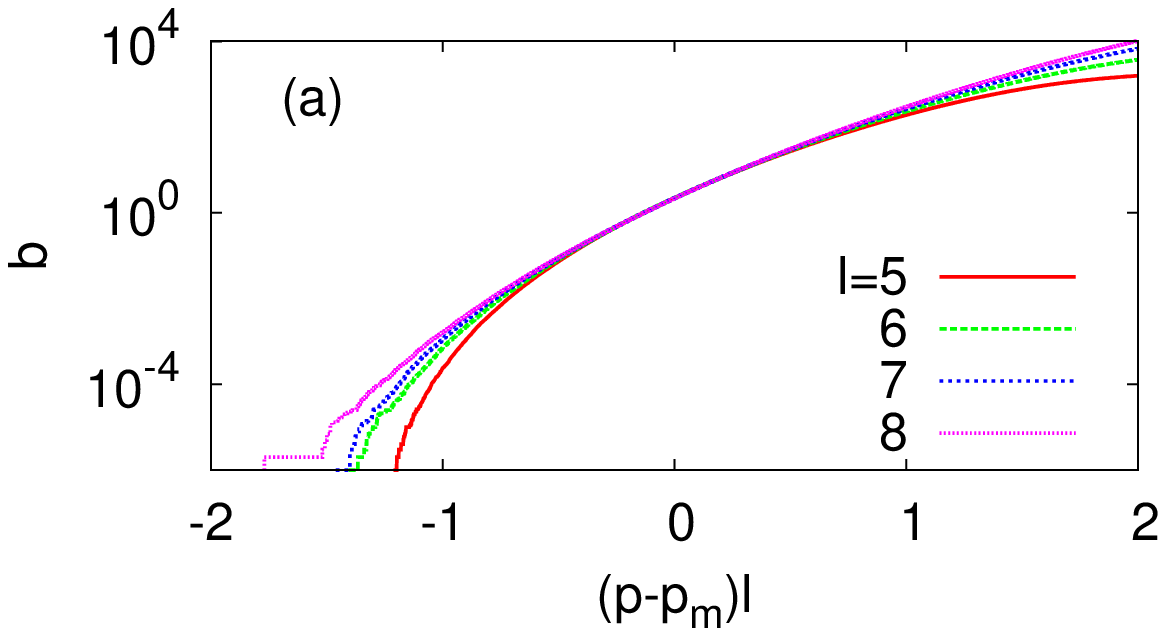} 
\includegraphics[width=0.45\textwidth]{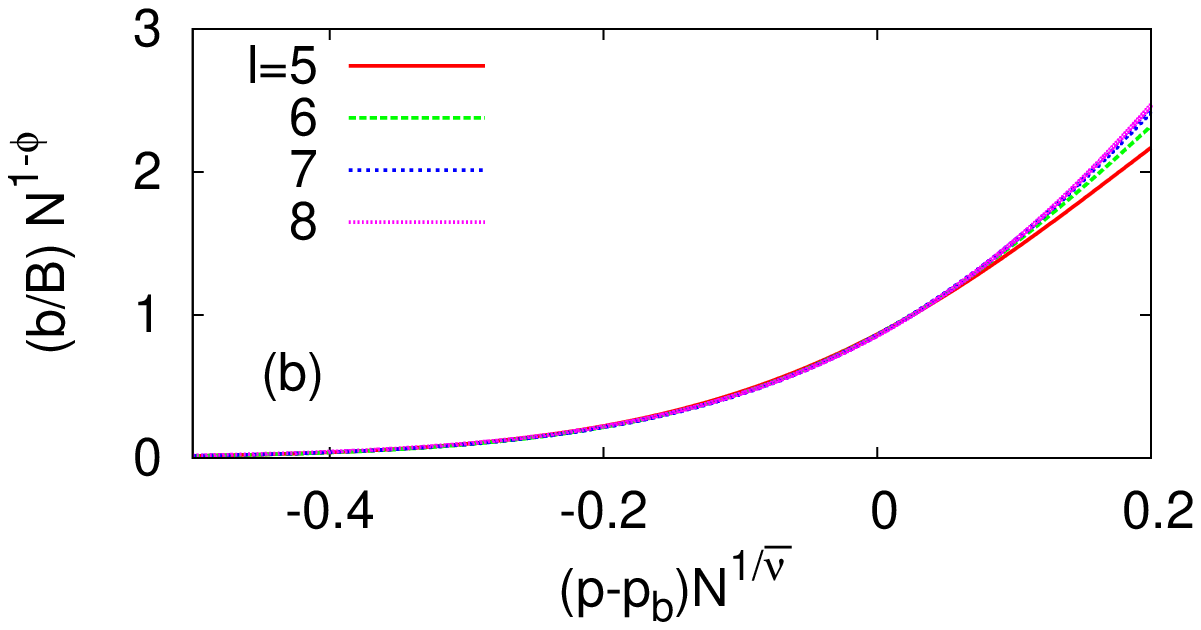} 
\caption{(Color online) Numerical results for $\{4,5\}$, averaged over
$10^6$ realizations. (a) Scaling plot with $p_m = 0.27$. 
(b) Scaling collapse with $p_b=0.52$, $\phi=0.82$, and $1/\bar\nu = 0.11$.}
\label{fig:4_5}
\end{figure}

\begin{figure}[tbp]
\includegraphics[width=0.45\textwidth]{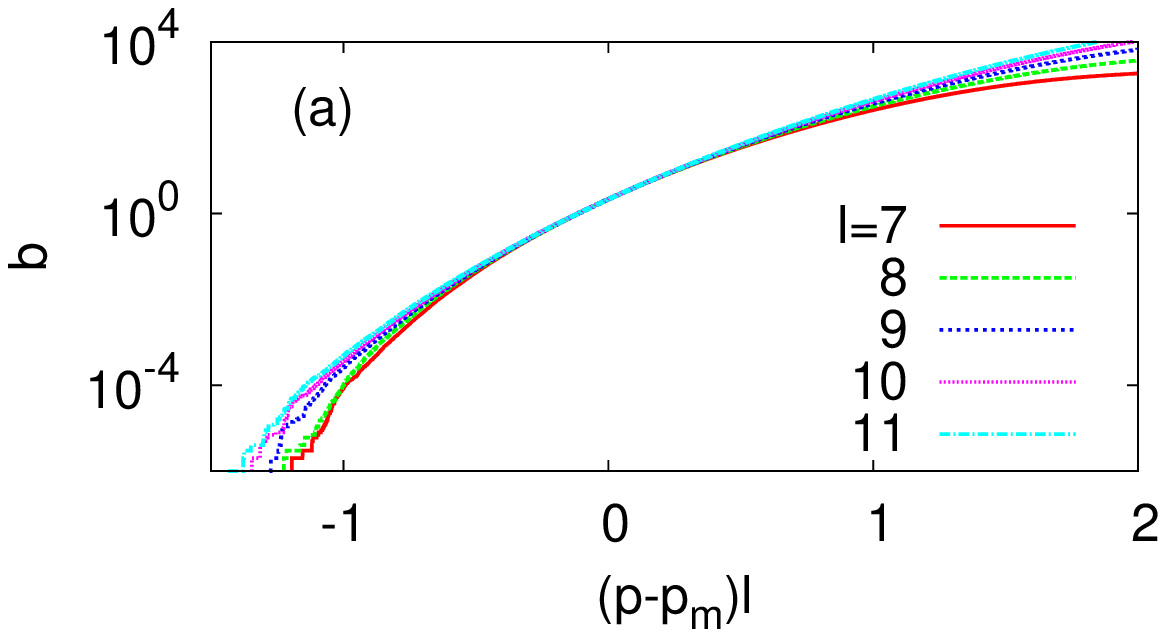} 
\includegraphics[width=0.45\textwidth]{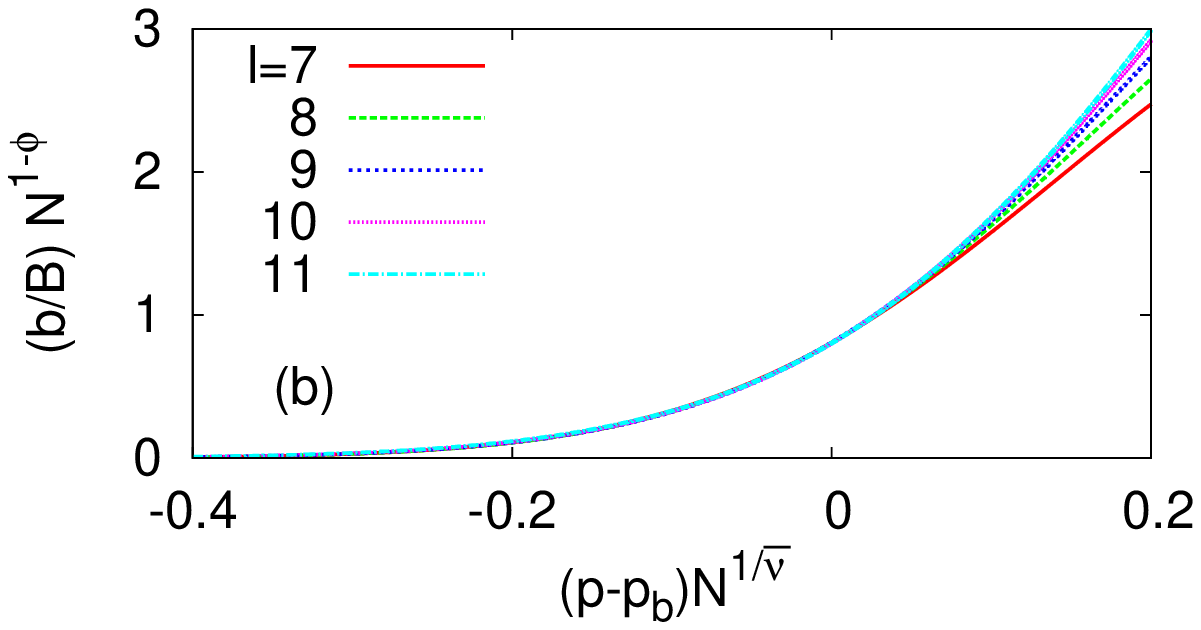} 
\caption{(Color online) Numerical results for $\{3,7\}$, averaged over
$10^6$ realizations. Here are shown the scaling plots (a) at $p_m = 0.20$
and (b) at $p_b=0.37$ with $\phi=0.82$ and $1/\bar\nu = 0.11$.}
\label{fig:3_7}
\end{figure}

In order to investigate the generality of our results, we also study two
additional hyperbolic lattices with different structures $\{m,n\}$. In all
cases we find precisely the same scenario. For the hyperbolic lattice
$\{4,5\}$, we find $p_m \approx 0.27$, $p_u = p_b \approx 0.52$
[Figs.~\ref{fig:4_5}(a) and \ref{fig:4_5}(b)].
The scaling behavior at the lower threshold $p_m$ is the same as for the
Cayley tree and for the lattice $\{7,3\}$. At the upper threshold $p_b=0.52$,
the critical indices $\phi=0.82$ and $1/\nu=0.11$ was determined from the
scaling collapse [Fig.~\ref{fig:4_5}(c)]. These values are identical to the
ones found for lattice $\{7,3\}$ within numerical accuracy.

For the lattice $\{3,7\}$ which is dual to $\{7,3\}$ the same agreement is
found:
The same size scaling is found at the lower threshold $p_m \approx
0.20$ [Fig.~\ref{fig:3_7}(a)]. At the upper one, $p_u \approx 0.37$,
the critical indices $\phi=0.82$ and $1/\bar\nu=0.11$ are found
[Fig.~\ref{fig:3_7}(b)].
This is again in striking agreement with the lattices $\{7,3\}$ and $\{4,5\}$.
Thus our results are consistent with a universal critical behavior at the
second threshold for all hyperbolic lattices $\{m,n\}$ provided both $m$ and $n$
are finite numbers. This critical behavior is distinct from the tree case
$\{\infty,n\}$, which has $\phi=1$ at $p=p_b$.
The present accuracy suggests that the critical indices are to good
approximation $\phi \approx 0.82$ and $1/\bar\nu \approx 0.11 \pm 0.01$.

We also note that while the lower threshold $p_m$ is still close to the tree
result $1/(n-1) = 1/4$ for $\{4,5\}$, the deviation becomes large for
$\{3,7\}$, and that the estimate $p_b = p_u = m/(m+n)$ in general only
gives a very crude estimate of the upper threshold.

\section{Summary}
\label{sec:summary}

We have investigated the percolation thresholds and the critical properties
of the percolation transitions for hyperbolic lattices using finite-size
scaling methods. 
Two distinct percolation thresholds were
found: The lower one corresponds to the threshold when the probability of
finding a cluster from the midpoint to the boundary becomes finite and the
second when the cluster containing the midpoint, with finite probability,
also contains a finite fraction of the boundary. This is in contrast to the
planar lattices which only possess a single percolation threshold
because the two thresholds above coincide. The Cayley tree was used as a
benchmark. It was found that the lower threshold for the hyperbolic lattices
has the same scaling properties as for the Cayley tree and that the
power-law dependencies characterizing the region between the two thresholds
are also like Cayley trees.
However, the second higher threshold has a different critical behavior. Our
results are consistent with a universal behavior at the higher threshold
for all hyperbolic lattices $\{m,n\}$ with $m$ and $n$ finite. This critical
behavior is characterized by two critical indices $\phi \approx 0.82$ and
$\nu \approx 0.11$. What actually determines these critical indices is still
an open question and will be the subject of future research. 

\acknowledgments
We are grateful to Bo S\"oderberg, Jae Dong Noh, Sang-Woo Kim, Hiroyuki
Shima, and Okyu Kwon for their inspirations and help.
S.K.B. and P.M. acknowledge the support from the Swedish Research Council
with the Grant No. 621-2002-4135. B.J.K. was supported by
the Korea Research Foundation Grant funded by the Korean
Government (MOEHRD) with Grant No. KRF-2007-313-C00282.
This research was conducted using the resources of High Performance
Computing Center North (HPC2N).

\end{document}